\documentclass [12pt]{article}

\usepackage{bm}
\usepackage{float}
\restylefloat{table}
\restylefloat{figure}
\usepackage{shuffle}

\usepackage{graphicx,amssymb,amsbsy,amsfonts,amssymb,amsmath}
\usepackage[usenames, dvipsnames]{xcolor}
\usepackage{calc}
\usepackage{url}

\allowdisplaybreaks
\usepackage{array}
\usepackage{bbm}
\usepackage{slashed}

\numberwithin{equation}{section}

\newif\ifdraft
\drafttrue
\newif\ifpreprint
\preprinttrue

\restylefloat{figure}
\allowdisplaybreaks

\newcommand{\req}[1]{(\ref{#1})}
\def\vev#1{\langle #1 \rangle}

\def\fc#1#2{\frac{#1}{#2}}
\def\h{\frac{1}{2}}
\newcommand{\nwc}{\newcommand}
\nwc{\ba}  {\begin{array}}
\nwc{\ea}  {\end{array}}
\nwc{\bdm} {\begin{displaymath}}
\nwc{\edm} {\end{displaymath}}

\nwc{\bea} {\begin{equation}\ba{lcl}}
\nwc{\eea} {\ea\end{equation}}

\nwc{\be} {\begin{equation}}
\nwc{\ee} {\end{equation}}

\nwc{\bda} {\bdm\ba{lcl}}
\nwc{\eda} {\ea\edm}

\nwc{\bc}  {\begin{center}}
\nwc{\ec}  {\end{center}}

\nwc{\ds}  {\displaystyle}

\nwc{\nn} {\nonumber}
\nwc{\nnn} {\nonumber \vspace{.2cm} \\ }
\nwc{\ra}{\rightarrow}
\nwc{\lra}{\longrightarrow}
\def\lf{\left}\def\ri{\right}

\nwc{\p} {\partial}
\nwc{\Tr}{{\rm Tr}}
\def\IR{{\bf R}}

\def\th{\theta}
\def\F{{_{q+1}F_q}}
\def\eps{\epsilon}
\def\z{\zeta}

\begin{document}

\title{\textbf{A Feynman Integral and its\\ Recurrences and Associators}  \\ }

\medskip
\author{\large Georg Puhlf\"urst and  
Stephan Stieberger\\[2cm]}
\date{}

\maketitle

\vspace{-2.5cm}
\centerline{\it  Max--Planck--Institut f\"ur Physik}
\centerline{\it Werner--Heisenberg--Institut}
\centerline{\it 80805 M\"unchen, Germany}

\vspace{3cm}
\begin{abstract}
We determine closed and compact expressions for the $\epsilon$--expansion of certain Gaussian hypergeometric functions expanded around half--integer values by explicitly solving for their recurrence relations. This $\epsilon$--expansion is identified with the normalized solution
of the underlying Fuchs system of four regular singular points.
We compute its  regularized zeta series (giving rise to two independent associators) whose ratio gives the  $\epsilon$--expansion at a specific value.
Furthermore, we use the well known one--loop massive bubble integral as an example to demonstrate how to obtain all--order $\epsilon$--expansions for Feynman integrals and how to construct representations for Feynman integrals in terms of generalized hypergeometric functions. We use the method of differential equations in combination with the recently established general solution for recurrence relations with non--commutative coefficients.

\end{abstract}
\bigskip\bigskip
\medskip
\vspace{-0.35cm}

\begin{flushright}
{\small  MPP--2015--259}
\end{flushright}


\thispagestyle{empty}

\newpage
\setcounter{tocdepth}{2}
\tableofcontents
\break

\section{Introduction}

Scattering amplitudes describe the interactions of physical states and play an important role to determine physical observables measurable at colliders. 
In perturbation theory at each order in the expansion scattering  amplitudes are comprised by a sum over Feynman diagrams 
with a fixed number of loops. Each individual Feynman diagram is represented by integrals over loop momenta and integrates to  
 functions, which typically depend on the Lorentz--invariant quantities of the external particles like  their momenta, masses
and scales. These functions are generically neither rational nor algebraic but give rise to a 
branch cut structure following from unitarity and the fact that virtual particles may go on-shell.

The class of functions describing Feynman integrals
are iterated integrals, elliptic functions and perhaps generalizations thereof.
To obtain physical results one is interested in their Laurent series expansion 
($\eps$--expansion)  about the integer value of the space--time dimension $D$ (typically $D =4-2\eps$). In the parameter space of the underlying higher transcendental functions this gives rise to an expansion 
w.r.t. small parameter $\eps$ 
around some fixed numbers, which may be integer or rational numbers. 
Expansions around integer values is in general sufficient for the evaluation of loop integrals arising in massless quantum field theories. However, the inclusion of particle masses in loop integrals or the evaluation of phase space integrals may give rise to half--integer values, cf. e.g. \cite{Weinzierl:2004bn}.

The module of hypergeometric functions \cite{Slater} is ubiquitous both in computing tree--level string amplitudes~\cite{Oprisa:2005wu} and in the evaluation of Feynman diagrams with loops cf. e.g. \cite{Smirnov}.
Therefore, finding an efficient procedure to determine power series expansion of these functions is an important problem.  Their underlying higher order differential equations lead to recurrence relations, which can be  solved  explicitly by the methods recently proposed in \cite{StSt}. This procedure gives an explicit solution to the recurrence relation providing 
for each order in  $\epsilon$ a closed, compact and analytic expression. This way we get  hands on the  $\eps$--expansions of this large family of functions \cite{StSt}.
For a subclass of the latter the coefficients of their expansions generically represent
multiple polylogarithms (MPLs), which give rise to periods of mixed Tate motives in algebraic geometry. Then, expansions around rational numbers $p/q$ naturally yield $q$--th roots of unity $\exp(2\pi ip/q)$ in the arguments of the MPLs
 \cite{Broadhurst:1998rz,Kalmykov:2008ge}.

Amplitudes in field--theory  very often can be described by certain differential equations 
or systems thereof with a given initial value problem subject to physical conditions 
\cite{Kotikov:1991pm}.
For generic parameters the corresponding differential equations for generalized hypergeometric functions are Fuchsian differential equations with the regular singular points at $0,1$ and $\infty$.
For a specific subclass (of at least ${}_2F_1$ hypergeometric functions to be specified later) at rational values $p/q$ of parameters (real parameters shifted by $p/q$) after a suitable coordinate transformation the underlying first order differential equations become a Fuchs system with $q+2$ 
regular singular points at $0,\exp(2\pi ir/q)$ and~$\infty$ with 
$r=1,\ldots,q$ (Schlesinger system).
By properly assigning the Lie algebra and monodromy representations of this linear system of differential equations their underlying fundamental solutions can be matched with the $\eps$--expansion  of specific  generalized hypergeometric functions. As a consequence, each order in 
$\eps$ is given by some combinations of MPLs and group--like matrix products carrying the 
information on the parameter.  
Furthermore, at special values the latter can be given in terms  
of  their underlying  regularized zeta series. The latter give rise to  
$q$ independent associators, 
which   are  defined as ratio of two solutions of the specific differential equation.
This way one obtains a very elegant way of casting the 
full $\epsilon$--expansion of certain generalized hypergeometric functions into the form dictated by the underlying Lie algebra structure and the analytic structure of MPLs.
Computing higher orders in the $\eps$--expansion is then reduced to simple matrix multiplications.
In this work we explicitly work out the case $q=2$, which is relevant to a Feynman integral to be discussed in this work and comment on the generic case
$q\neq 2$.

Feynman integrals are classified according to their topologies. A topology includes all integrals, that consist of the same set of propagators but have different powers thereof. There are integration--by--parts (IBP) identities \cite{Chetyrkin:1981qh}, which allow to reduce all integrals of a topology to a set of master integrals (MIs). One way to compute MIs is to apply the method of differential equations \cite{Kotikov:1991pm} (for reviews, see \cite{Argeri:2007up}).
The idea of this method is to take derivatives of MIs w.r.t. kinematic invariants and masses. The results are combinations of integrals of the same topology, which can again be written in terms of MIs using the IBP reduction. This way one obtains differential equations for the MIs. In practice the goal is to solve these equations in a Laurent expansion around $\epsilon=0$. If the differential equations take a suitable form the coefficient functions of the $\epsilon$--expansion can be given in a straightforward iterative form \cite{Henn:2013pwa}. By replacing integrations with integral operators the iterative solutions can be written as recurrence relations with non--commutative coefficients. Solving these recurrences yields the all--order 
$\epsilon$--expansions for Feynman integrals. This approach allows to give the Laurent expansions as infinite series explicitly in terms of iterated integrals, thereby providing solutions, 
which are exact in $\epsilon$. This outcome is equivalent to the representation generic to hypergeometric functions and their generalization, with the advantage that the behaviour of the Feynman integral at $\eps=0$ can be extracted directly. 

The present work  is organized as follows. In section 2 we show how to systematically and 
efficiently derive $\epsilon$--expansions of 
hypergeometric function $\scriptstyle{_2F_1\lf[{\eps a,\eps b\atop \eps c+\h};z\ri]}$ with one half--integer 
parameter. 
In section 2.1 we introduce harmonic polylogarithms (HPLs), integral operators and related objects. In section 2.2 we discuss 
the differential equation satisfied by the hypergeometric function $\scriptstyle{_2F_1\lf[{\eps a,\eps b\atop \eps c+\h};z\ri]}$. A given order of its $\eps$--expansion can be derived from this differential 
equation based on 
computing successively all lower orders. 
Next, in section 2.3 
by transforming the differential equation to recurrence relations 
we derive the terms in the power series solution for any  given order in $\eps$.   This strategy yields an all--order expansion, 
i.e. an infinite 
Laurent series in $\epsilon$ explicitly in terms of iterated integrals. In section 2.4 we 
investigate the underlying Fuchs system involving four regular singular points describing the hypergeometric function $\scriptstyle{_2F_1\lf[{a,b\atop c+\h};z\ri]}$ with one half--integer parameter. For the latter we derive analytic solutions 
in terms of hyperlogarithms and group--like matrix elements encoding the information on the parameters 
$a,b$ and $c$. One of these solutions will be matched 
with the relevant $\epsilon$--expansion of 
  the hypergeometric function $\scriptstyle{_2F_1\lf[{\eps a,\eps b\atop \eps c+\h};z\ri]}$. 
As a consequence the coefficients of the $\eps^k$--order in the power series expansion  are given by a set of the MPLs of  degree $k$ entering the fundamental solution  supplemented 
by matrix products of $k$ matrices.  
Furthermore, we construct the two regularized zeta series generic to the underlying Fuchs system.
The latter gives rise to two independent associators whose ratio will be related to the hypergeometric function  $\scriptstyle{_2F_1\lf[{a,b\atop c+\h};1\ri]}$  at the special point $z=1$.
In section 3 as an example 
we discuss 
the $\epsilon$--expansion of Feynman integrals using a massive one--loop integral. 
The all--order expansion is derived via the method of differential equations in section 3.1 and 
this result is used to construct a representation in terms of a hypergeometric function in section 3.2. 
Benefiting from the fact that the all--order result is exact in $\epsilon$, we can give the hypergeometric representation of the Feynman integral not only in $D=4-2\epsilon$ dimensions but also in general dimensions $D$.
Finally, in section 4 we present  for the hypergeometric function 
$\scriptstyle{_2F_1\lf[{a,b\atop 1-\fc{p}{q}+c};z\ri]}$
the underlying Fuchs system involving $q+2$ regular singular points.  We comment on its generic solutions and the underlying $q$ associators.

\section{Hypergeometric Function with half--integer parameters}

The Gaussian hypergeometric function ${}_2F_1$ is given by the power series \cite{Slater}
\begin{align}\label{Hyper}
{}_2F_{1}\left[{a,b\atop c};z\right]=\sum_{m=0}^\infty\ 
\frac{(a)^m(b)^m}{(c)^m}\ \frac{z^m}{m!}\ ,
\end{align}
with parameters $a,b,c\in\IR$ and with the Pochhammer (rising factorial) symbol:
$$(a)^n=\frac{\Gamma(a+n)}{\Gamma(a)}
\ .$$
The series \req{Hyper} converges absolutely at the unit circle $|z| = 1$ if the parameters meet the following condition:
\begin{align}\label{conv}
c-a-b>0\ .
\end{align}
In the sequel we want to investigate hypergeometric functions \req{Hyper} with some of their  parameters shifted by $1/2$. This gives rise to the following five possibilities
\be\label{possa}
{}_2F_{1}\left[{a,b\atop \h+c};z\ri],\ {}_2F_{1}\left[{\h+a,\h+b\atop \h+c};z\ri],\ {}_2F_{1}\left[{\h+a,b\atop \h+c};z\ri],\ {}_2F_{1}\left[{\h+a,b\atop 1+c};z\ri]\ ,
\ee
and:
\be\label{possb}
{}_2F_{1}\left[{\h+a,\h+b\atop 1+c};z\ri]\ .
\ee
It has been shown in \cite{Kalmykov:2006pu} that 
the class of four hypergeometric functions \req{possa} can algebraically be written in terms of only one, e.g.:
$${}_2F_{1}\left[{a,b\atop \frac12+c};z\ri]\ .$$
If the functions  \req{possa} (with the same argument $z$)  have in their parameters additional integer shifts, they can be  expressed in terms of \req{possa} and first derivatives thereof  and some polynomials in the parameters $a,b,c$ and $z$.
Therefore, in this section we shall apply our new technique \cite{StSt} to solve for 
recurrences to compute the following  $\epsilon$--expansion: 
\begin{align}\label{3hg}
{}_2F_{1}\left[{a\epsilon,b\epsilon\atop \h+c\epsilon};z\ri]=
\sum\limits_{k}\epsilon^k\ u_{k}(z)\ .
\end{align}
The coefficient functions $u_k(z)$ are expressible in terms of harmonic polylogarithms (HPLs) with rational coefficients \cite{Kalmykov:2006hu}.
On the other hand, the type \req{possb} involves elliptic functions and will not be discussed here. E.g. we have:
${}_2F_{1}\left[{\h,\h\atop 1};z\ri]=\tfrac{2}{\pi}\ K(\sqrt z)$ 
with $K$ the elliptic function  of first kind.

\subsection{Harmonic polylogarithms, integral operators and the generalized operator product}
\label{Section21}

HPLs of weight $w\geq2$ are defined recursively as \cite{Remiddi}
\begin{align}
\begin{split}\label{hpl}
H(m_0,\vec{m};y)&=\int_0^ydt \;g(m_0;t) H(\vec{m};t)\ , \ \ m_i\in\{0,1,-1\}\ ,  \ \ (m_0,\vec{m})\neq(\underbrace{0,\ldots,0}_{w})\ , \\
H(\underbrace{0,\ldots,0}_{w};y)&=\frac{1}{w!}\ln^wy\ ,
\end{split}
\end{align}
with the multiple index $\vec{m}=(m_1,m_2,\ldots,m_{w-1})$ and
\begin{align}
g(0;y)=\frac{1}{y}\ , \ \
g(1;y)=\frac{1}{1-y}\ , \ \
g(-1;y)=\frac{1}{1+y}\ .
\end{align}
Weight $w=1$ functions are simple logarithms:
\begin{align}
H(0;y)&=\ln(y) \ , \ \ H(1;y)=-\ln(1-y) \ , \ \ H(-1;y)=\ln(1+y) \ .
\end{align}

In the next subsection we will encounter differential equations satisfied by functions for which we 
know their boundary  conditions at $y=1$. Therefore, it is advantageous to introduce the following 
integral operators
\begin{align}\label{3jops}
\begin{split}
J(0)f(y)&=\int_{1}^y\frac{dt}{t}f(t)\ ,\\[2mm]
J(1)f(y)&=\int_{1}^y\frac{dt}{1-t}f(t)\ ,\\[2mm]
J(-1)f(y)&=\int_{1}^y\frac{dt}{1+t}f(t)\ ,
\end{split}
\end{align}
with their integrations starting at the point 1. We shall 
use a shorter notation for products of these operators:
\begin{align}
J(m_1,m_2,\ldots,m_w)\equiv J(m_1)J(m_2)\ldots J(m_d)\ , \ \  m_i\in\{0,1,-1,\theta\}\ , \ \ J(\theta)\equiv y\frac{d}{dy}\ .
\end{align}
Products of integral operators \req{3jops} acting on functions independent of $y$, like the constant function\footnote{In the sequel we will not write this 1.} 1, can recursively be written as HPLs using
\begin{align}
\begin{split}\label{jtohpl}
J(m_1)\; 1&=H(m_1;y)-H(m_1;1)\ , \\
J(m_1)H(m_2,\ldots,m_w;y)&=H(m_1,m_2,\ldots,m_w;y)-H(m_1,m_2,\ldots,m_w;1)\ , \\
\end{split}
\end{align}
with $m_i\in\{0,1,-1\}$, e.g.:
\begin{align}
\begin{split}
J(\underbrace{0,\ldots,0}_{w})\;1&=H(\underbrace{0,\ldots,0}_{w};y)\ , \\
J(1,0)\;1&=H(1,0;y)-H(1,0;1)\ , \\
J(0,1,0)\;1&=H(0,1,0;y)-H(0,1,0;1)-H(0;y)H(1,0;1)\ , \\
J(0,-1,0)\;1&=H(0,-1,0;y)-H(0,-1,0;1)-H(0;y)H(-1,0;1)\ .
\end{split}
\end{align}
Since HPLs with argument 1 can have divergences, the second terms on the r.h.s. of eqs. \req{jtohpl} reveal that some products of integral operators are ill-defined. This happens when $J(1)$ is the rightmost operator of a product acting on a constant. It can be seen in the following sections that these cases never appear in our results. Divergent HPLs also arise for other operator products, however, they can be removed using the underlying product algebra. Consider for example $J(1,1,0)$ and its representation in terms HPLs following from eqs. \req{jtohpl}:
\begin{align}\label{div}
J(1,1,0)=H(1,1,0;y)-H(1,0;1)H(1;y)-H(1,1,0;1)+H(1,0;1)H(1;1)\ .
\end{align}
According to the identity
\begin{align}
H(1,0;1)H(1;1)=-H(0,1,1;1)+H(1,1,0;1)
\end{align} 
the divergences in the last two terms of \req{div} cancel each other.

The essential difference between the integral operators \req{3jops} and the integrations in the definition \req{hpl} of HPLs is the lower bound. See also eq. \req{Lappo} or Ref. \cite{Gon2001} for polylogarithms with lower integration limits not chosen to be zero.

Products of integral operators \req{3jops} are closely related to the `symbol' of the underlying function, together with the prescription that the integration path goes from 1 to z. Indeed, a sequence of integral operators immediately determines a `symbol', which when integrated against this path turns it into the Chen iterated integral representation of the function (see e.g. Ref. \cite{Duhr} for an introduction for physicists).

Furthermore we introduce the generalized operator product
\begin{align}\label{gop}
\{c_1^{j_1},c_2^{j_2},\ldots,c_n^{j_n}\}\ ,
\end{align}
as the sum of all the
\begin{align}
\binom{\sum\limits_{\alpha=1}^{n}j_\alpha}{j_1,j_2,\ldots,j_n}
\end{align}
possible distinct permutations of non--commutative factors $c_i$, each one appearing $j_i$ times ($i=1,2,\ldots,n$). We will call the non--negative integers $j_i$ indices and the factors $c_i$ arguments of the generalized operator product \req{gop}. Some examples are:
\begin{align}
\begin{split}
\{c_1^2,c_2\}&=c_1^2c_2+c_1c_2c_1+c_2c_1^2\ , \\
\{c_1,c_2,c_3\}&=c_1c_2c_3+c_1c_3c_2+c_2c_1c_3+c_2c_3c_1+c_3c_1c_2+c_3c_2c_1\ ,\\
\{c_1^{0},c_2^{j_2},\ldots,c_n^{j_n}\}&=\{c_2^{j_2},\ldots,c_n^{j_n}\}\ ,\\
\{c^j\}&=c^j\ ,\\
\{c^0\}&=1\ .
\end{split}
\end{align}
A recursive definition and basic properties of the generalized operator product can be found in \cite{StSt}.
The object \req{gop} is useful to handle non--commutative quantities, such as the integral operators \req{3jops}.

\subsection{Differential equations for hypergeometric functions}
In this section we describe some of the achievements originally developed in \cite{Kalmykov:2006hu} for the calculation of expansions \req{3hg}. Some of the formulas derived here are the foundation of the method we use in the next section to obtain the all--order expansion.

Applying the differential operator $\theta:=z\frac{d}{dz}$ on the series \req{Hyper}, it is easy to show, that hypergeometric functions satisfy:
\begin{align}
 \begin{split}
\label{3hgrel}
 \left (\theta + a \right ){}_2F_{1}\left[{a,b\atop c};z\right] &= a\;{}_2F_{1}\left[{1+a,b\atop c};z\right]\ ,\\[2mm]
  \left (\theta + b \right ){}_2F_{1}\left[{a,b\atop c};z\right] &= b\;{}_2F_{1}\left[{a,1+b\atop c};z\right]\ ,\\[2mm]
\left (\theta+ c - 1 \right ){}_2F_{1}\left[{a,b\atop c};z\right] &= (c - 1)\; {}_2F_{1}\left[{a,b\atop c-1};z\right]\ ,\\[2mm]
\theta\; {}_2F_{1}\left[{a,b\atop c};z\right] &= z\frac{ab}{c}
\; {}_2F_{1}\left[{1+a,1+b\atop1+ c};z\right]\ .
 \end{split}
\end{align}
Combining these relations yields the differential equation:
\begin{align}\label{3hgdgl}
z(\theta+a)(\theta+b)\;{}_2F_{1}\left[{a,b\atop c};z\right]=\theta\ (\theta+c-1)\ {}_2F_{1}\left[{a,b\atop c};z\right]\ .
\end{align}
For the hypergeometric function
\begin{align}
\varphi_1(z):={}_2F_{1}\left[{a\eps,b\eps\atop\frac12+ c\eps};z\right]
\end{align}
the second order differential equation \req{3hgdgl} in $z$ can be written as a system of two first order differential equations in the variable 
\begin{align}\label{subst}
y=\frac{\sqrt{z-1}-\sqrt{z}}{\sqrt{z-1}+\sqrt{z}}\ , 
\end{align}
for $\varphi_1(y)$ and $\varphi_2(y):=y\frac{d}{dy}\varphi_1(y)$:
\begin{align}
\begin{split}\label{syst0}
\frac{d}{dy}\varphi_1(y)&=\frac{1}{y}\varphi_2\ ,\\[2mm]
\frac{d}{dy}\varphi_2(y)&=\frac{2c\eps}{1-y}\varphi_2(y)+\frac{(a+b)\eps}{y}\varphi_2(y)-2\eps\frac{a+b-c}{1+y}\varphi_2(y)-\frac{a b \eps^2}{y}\varphi_1(y)\ .
\end{split}
\end{align}
Inserting the expansion \req{3hg} and using, that the resulting differential equations are valid at any order in $\epsilon$, yields iterative differential equations for the coefficient functions $u_k(y)$ and $v_k(y):=y\frac{d}{dy}u_k(y)$:
\begin{align}\begin{split}\label{syst}
\frac{d}{dy}u_k(y)&=\frac1y v_k(y)\ , \\[2mm]
\frac{d}{dy}v_k(y)&=\frac{2c}{1-y}v_{k-1}(y)+\frac{a+b}{y}v_{k-1}(y)-2\frac{a+b-c}{1+y}v_{k-1}(y)-\frac{a b}{y}u_{k-2}(y)\ .
\end{split}
\end{align}
Eq. \req{Hyper} can be used to determine boundary conditions at $z=0$. This point transforms to the point $y=1$ under \req{subst}, so that the boundary conditions are $v_k(1)=0$ for $k\geq 0$
and $u_k(y=1)=0$ for $k\geq 1$. The system \req{syst} can then be solved iteratively for $k\geq1$ via:
\begin{align}
\begin{split}\label{diffeq}
u_k(y)&=\int_{1}^{y}\frac{dt}{t}v_k(t)\  ,\\[2mm]
v_k(y)&=\int_{1}^{y}dt\left[\frac{2c}{1-t}+\frac{a+b}{t}-2\frac{a+b-c}{1+t}\right]v_{k-1}(t)-ab\int_{1}^y\frac{dt}{t}u_{k-2}(t)\ .
\end{split}
\end{align}
From eqs. \req{Hyper} and \req{conv} follow the lowest orders $u_0(z)=1$ and $u_{k}(z)=0$ for $k<0$. Starting with those we can use eqs. \req{diffeq} to straightforwardly calculate the $\epsilon$--expansion \req{3hg} order by order, e.g.:
\begin{align}\begin{split}
u_1(y)&=0\\
u_2(y)&=-ab\;J(0,0)=-ab\;H(0,0;y)\label{1stord}\\
u_3(y)&=-2abc\;J(0,1,0)+2ab(a+b-c)J(0,-1,0)-ab(a+b)J(0,0,0)\\
  &=-2abc\left[H(0, 1, 0; y) - H(0, 1, 0;1) - H(0;y) H(1, 0;1)\right]+2ab(a+b-c)\\
  &\times \left[H(0, -1, 0; y) - H(0, -1, 0;1) - H(0;y) H(-1, 0;1)\right]-ab(a+b)H(0, 0, 0; y)\\
u_4(y)&=-4 abc^2 J(0, 1, 1, 0) - 4ab (a + b - c)^2 J(0, -1, -1, 0)\\
 & - ab(a + b)^2 J(0, 0, 0, 0) + a^2 b^2 J(0, 0, 0, 0) + 4 a b c(a + b - c)  \\
 &\times\left[J(0, 1, -1, 0) + J(0, -1, 1, 0)\right] - 2 a b c (a + b) \left[J(0, 1, 0, 0) + J(0, 0, 1, 0)\right] \\
&+ 2 a b (a + b) (a+ b - c) \left[J(0, -1, 0, 0) + J(0, 0, -1, 0)\right]  
\end{split}\end{align}
We used eqs. \req{jtohpl} to give $u_2(y)$ and $u_3(y)$ in terms of HPLs. This can be achieved for $u_4(y)$ and higher order coefficient functions as well.

The iterative computation of Laurent expansions of hypergeometric functions described in this section has first been presented in \cite{Kalmykov:2006hu}. It has been applied to similar function, e.g. to generalized hypergeometric functions with integer parameters in \cite{Kalmykov}. Eqs. \req{diffeq} are the main results of this subsection. They allow to straightforwardly calculate the expansion \req{3hg} up to any order in $\epsilon$. However, is is not possible to obtain a given order without knowing its lower orders. 
This problem will be solved in the next section. 

\subsection{Recurrence relations for generalized hypergeometric functions}

In this section we present the all--order expansions for the hypergeometric functions \req{3hg}. The idea is to write the differential equations for the coefficient functions as recurrence relations\footnote{Similar manipulations however in a different context are used in \cite{Ablinger:2015tua} for the calculation of Feynman integrals.}. This is achieved by replacing the derivatives and integrations in the iterative solutions of these differential equations with differential and integral operators, respectively. In \cite{Boels} such recurrence relations have been used to calculate $\alpha'$--expansions for generalized hypergeometric functions, which enter open string amplitudes. The recurrence relations are homogeneous and linear. Their coefficients, which involve integral- and differential-operators, are non--commutative. The iterative solutions \req{diffeq} of the differential equations \req{syst} are actually equivalent to the recurrence relations. With the latter it becomes, however, more obvious how to calculate Laurent expansions order by order.

More importantly, the general solution for this type of recurrence relations has been presented in \cite{StSt}. Hence the all--order expansion of \req{3hg} can now systematically be constructed and 
straightforwardly be given in a closed form. By \textit{all--order} we mean a representations for the coefficient functions $u_{k}$, which include $k$ as a variable and therefore hold for all orders in $\epsilon$. In contrast to that, the method of the previous section allows to compute $\epsilon$--expansions order by order starting with $u_{0}$, $u_{1}$, $u_{2}$ and so on. 
In other words the formula for $u_{k}$, no matter if in the form of a iterative solution to 
differential equations as discussed in the previous section or as a recurrence relation as presented in the following, is not given in terms of HPLs or similar functions. Instead, coefficient functions of lower orders are included. On the other hand our all--order result obtained from the solution of the recurrence relation gives coefficient functions for all orders \textit{explicitly} in terms of iterated integrals.

Combining eqs. \req{diffeq} yields the second order recurrence relation
\begin{align}\label{recrelJ}
u_k(y)=c_1u_{k-1}(y)+c_2u_{k-2}(y)\ , \ \ k\geq 2 \ ,
\end{align}
with the non--commutative coefficients
\begin{align}
\begin{split}
c_1&=2c\;J(0,1,\theta)-2(a+b-c)J(0,-1,\theta)+(a+b)J(0)\ , \\
c_2&=-ab\;J(0,0) \ .
\end{split}
\end{align}
With the initial values $u_0(y)=1$ and $u_1(y)=0$ the solution of eq. \req{recrelJ} reads
\begin{multline}\label{3half}
u_k(y)=\sum\limits_{\substack{l_1+l_2+l_3+2m\\=k-2}}(-1)^{l_2+m+1}2^{l_1+l_2}c^{l_1}(a+b-c)^{l_2}(a+b)^{l_3}(ab)^{m+1}\\
\times J(0)\{J(1)^{l_1},J(-1)^{l_2},J(0)^{l_3},J(0,0)^{m}\}J(0)\ .
\end{multline}
Let us demonstrate that, in contrast to the findings in \cite{Kalmykov:2006hu}, this result allows to obtain any order of the expansions \req{3hg} directly without using lower orders.
For instance $k=3$ in \req{3half} yields the condition $l_1+l_2+l_3+2m=1$ for the sum over non--negative integers $l_1,l_2,l_3$ and $m$. This equation has three solutions: $(l_1,l_2,l_3,m)=(1,0,0,0)$, $(0,1,0,0)$ and $(0,0,1,0)$. It is easy to check, that they give
 $$-2abcJ(0,1,0)\ ,\ \ 2ab(a+b-c)J(0,-1,0) \text{ and } -ab(a+b)J(0,0,0)\ ,$$ respectively. This is in accordance with the expression for $u_3(y)$ given in \req{1stord}. Higher orders can be evaluated the same way, e.g. for $k=5$ there are 13 terms with the following summation indices:
\begin{align*}
  (l_1,l_2,l_3,m)=&(3,0,0,0),~(0,3,0,0),~(0,0,3,0),~(1,0,0,1),~(0,1,0,1),~(0,0,1,1),~(1,1,1,0)\\
  &(2,1,0,0),~(2,0,1,0),~(1,2,0,0),~(0,2,1,0),~(1,0,2,0)\text{ and }(0,1,2,0)\ .
\end{align*}
Every summand can be calculated straightforwardly, e.g. for $(l_1,l_2,l_3,m)=(0,0,1,1)$ we get $-2a^2b^2(a+b)J(0,0,0,0,0)$.

According to the discussion at the beginning of this section a variety of other hypergeometric 
functions ${}_2F_1$ with parameters, which differ from those in \req{3hg} by integer or half--integer 
shifts, can algebraically be expressed as linear combinations of the function \req{3hg} and derivatives thereof.
Thus our result \req{3half} allow to construct the all--order expansions for these functions as well. 
One of these functions appears in section 3 in a representation of a Feynman integral.

\subsection{Fuchsian system and its associators}
\def\Li{{\cal L}i}
\def\no{\nonumber}
\def\z{\zeta}

The differential equation \req{3hgdgl} 
\begin{align}\label{3hgdglnew}
z(\theta+a)(\theta+b)\;{}_2F_{1}\left[{a,b\atop \h+c};z\right]=\theta\ \left(\theta+c-\h\right)\ {}_2F_{1}\left[{a,b\atop \h+c};z\right]\ .
\end{align}
for the function
\be
{}_2F_{1}\left[{a,b \atop \h+c};z \right]\label{Start}
\ee
can be written as a system of two first order differential equations of Fuchsian class in the variable $y$ defined  in \req{subst}
\be\label{Schlesinger}
\fc{d\Phi}{dy}=\lf(\sum_{i=0}^2\fc{A_i}{y-y_i}\ri)\ \Phi\ ,
\ee
with the singularities $y_0=0,\ y_1=1,\ y_2=-1$, the vector
$$
\Phi=\lf({\varphi_1\atop \varphi_2}\ri)\ ,
$$
and (cf. also \req{syst0}): 
\bea
\varphi_1(y)&:=&\ds{{}_2F_{1}\left[{a,b \atop \h+c};z \right]\equiv u(z)\ ,}\\
\varphi_2(y)&:=&\ds{y\fc{d}{dy}\ \varphi_1\equiv v(z)\ .}
\eea
In \req{Schlesinger} the $2\times2$ matrices $A_i$ are given by:
\be\label{matrices}
A_0=\begin{pmatrix}
0&1\\
-ab&a+b
\end{pmatrix}\ \ \ ,\ \ \ A_1=\begin{pmatrix}
0&0\\
0&-2c
\end{pmatrix}\ \ \ ,\ \ \ A_2=\begin{pmatrix}
0&0\\
0&2c-2a-2b
\end{pmatrix}\ .
\ee
The system \req{Schlesinger} has four regular singular points at $y=0,1,-1$ and $y=\infty$
and is known as Schlesinger system. 
Recently, in Ref. \cite{StSt} we have thoroughly discussed 
the latter and its generic solution given in terms of hyperlogarithms ~\cite{Goncharov}.

Hyperlogarithms are defined recursively from words $w$ built from an alphabet $\{w_0,w_1,\ldots\}$
 (with $w_i\simeq A_i$) with $l+1$ letters:
\bea\label{hyperlog}
\ds{L_{w_0^n}(x)}&:=&\ds{\fc{1}{n!}\ \ln^n (x-x_0)\ ,\ n\in {\bf N}\ ,}\\[5mm]
\ds{L_{w_i^n}(x)}&:=&\ds{\fc{1}{n!}\ \ln^n\lf(\fc{x-x_i}{x_0-x_i}\ri)\ ,\ 1\leq i\leq l\ ,}\\[5mm] 
\ds{L_{w_iw}(x)}&:=&\ds{\int_0^x\fc{dt}{t-x_i}\ L_w(t)\ \ \ ,\ \ \ L_1(x)=1\ .}
\eea
Generically, we have
\be
L_{w_0^{n_l-1}w_{\sigma_l}\ldots w_0^{n_2-1}w_{\sigma_2}w_0^{n_1-1}w_{\sigma_1}}(x)=
(-1)^l\ {\cal L}i_{n_l,\ldots,n_1}\lf(\fc{x-x_0}{x_{\sigma_l}-x_0},
\fc{x_{\sigma_l}-x_0}{x_{\sigma_{l-1}}-x_0},\ldots
\fc{x_{\sigma_3}-x_0}{x_{\sigma_2}-x_0},
\fc{x_{\sigma_2}-x_0}{x_{\sigma_1}-x_0}\ri)\ ,
\ee
with the MPLs:
\be
{\cal L}i_{n_1,\ldots,n_l}(z_1,\ldots,z_l)=\sum_{0<k_l<\ldots <k_1}
\fc{z_1^{k_1}\cdot\ldots\cdot z_l^{k_l}}{k_1^{n_1}\cdot\ldots\cdot k_l^{n_l}}\ .
\ee
More generally one considers the iterated integrals \cite{LD}
\be\label{Lappo}
L^b_{w_iw}(x)=\int_b^x\fc{dt}{t-x_i}\ L^b_w(t)\ ,
\ee
with some base point $b$, which can be entirely expressed in terms of the objects \req{hyperlog}.
Of course, we have $L_w^0\equiv L_w$.
In the case $l=2$ under consideration  with the three singular points $x_0=0,\ x_1=1$ and $x_2=-1$ the alphabet $A$ consists of three letters $A^\ast=\{w_0,w_1,w_2\}$ and is directly related to the differential forms 
$$\fc{dx}{x}\ ,\ \fc{dx}{x-1}\ ,\ \fc{dx}{x+1}$$ 
appearing in \req{Schlesinger}. The corresponding hyperlogarithms \req{hyperlog} boil down to the 
HPLs discussed in section \ref{Section21}.

A group-like solution $\Phi$ to \req{Schlesinger}  taking values in ${\bf C}\vev{A}$  with the alphabet 
$A=\{A_0,A_1,A_2\}$ can be given as formal weighted sum over iterated integrals  (with the weight given by the number of iterated integrations)
\be\label{genSOL}
\Phi(x)=\sum_{w\in A^\ast}L_w(x)\ w\ ,
\ee
with the MPLs \req{hyperlog}. The solution \req{genSOL} can be constructed recursively and built by 
Picard's iterative methods\footnote{The non--vanishing entries of the matrices \req{matrices} are homogeneous polynomials of a given degree in the non--vanishing parameters $a,b,c$. 
These degrees are in one--to--one correspondence with the  number of derivatives $y\tfrac{d}{dy}$ in  the matrix equation \req{Schlesinger} allowing for an iteration \req{genSOL}
 w.r.t. the length of the word $w\in A^\ast$.}.
It is not possible to find power series solutions in $x$ expanded at $x=0,1$ or $x=-1$, because they have essential singularities at these points.
However, one can construct unique analytic solutions $\Phi_i$ normalized at $x=x_i$ with the asymptotic behaviour $x\ra x_i$ as
\be
\Phi_i(x)\lra (x-x_i)^{A_i}\ \ \ ,\ \ \ i=0,1,2\ , 
\ee
respectively. 
Generalizing \req{genSOL} to
\be\label{genSOLL}
\Phi_i(x)=\sum_{w\in A^\ast}L^i_w(x)\ w\ ,
\ee
we arrive at the following expansions\footnote{To compute the relevant iterated integrals \req{Lappo}
we have profited from \cite{Panzer}.} 
\begin{align}
\Phi_0(x)&=1+A_0\ \ln x+A_1\ \ln(1-x)+A_2\ \ln(x+1)-A_0A_1\ \Li_2(x)-A_0A_2\ \Li_2(-x)\no\\
&+A_1A_0\ [\ \ln x \ln(1-x)+\Li_2(x)\ ]+A_2A_0\ [\ \ln x\ \ln(1+x)+\Li_2(-x)\ ]\no\\
&+A_1A_2\ \Li_{1,1}(x,-1)+A_2A_1\ \Li_{1,1}(-x,-1)+\h\ \ln^2x\ A_0^2+\h\ \ln^2(1-x)\ A_1^2\no\\
&+\h\ \ln^2(1+x)\ A_2^2+\ldots\ ,\label{PHI0}
\end{align}
\begin{align}
\Phi_1(x)&=1+A_0\ \ln x+A_1\ \ln(x-1)+A_2\ [\ -\ln 2+\ln(x+1)\ ]\no\\
&+A_0A_1\ [\ \zeta_2-\Li_2(x)+i\pi \delta_x\ \ln x\ ]
+A_1A_0\ [\ -\zeta_2+\Li_2(x)+\ln x \ln(1-x)\ ]\no\\
&+A_0A_2\ [\ -\h\zeta_2-\Li_2(-x)-\ln 2\ln x\ ]+
A_2A_0\ [\ \h\zeta_2+\Li_2(-x)+\ln x\ \ln(1+x)\ ]\no\\
&+A_1A_2\ [\ -\h\zeta_2+\h\ln^22-\ln2\ln(1-x)+\Li_{1,1}(x,-1)\ ]\no\\
&+A_2A_1\ [\ \h\zeta_2-\h\ln^22+\Li_{1,1}(-x,-1)-i\pi \delta_x\ \{\ln2-\ln(1+x)\}\ ]\no\\
&+\h\ \ln^2x\ A_0^2+\h\ \ln^2(x-1)\ A_1^2+\h\ [\ln2-\ln(1+x)]^2\ A_2^2
+\ldots\ ,\label{PHI1}
\end{align}
and:
\begin{align}
\Phi_{2}(x)&=1+A_0\ \ln (-x)+A_1\ [\ \ln(1-x)-\ln 2\ ]+A_2\ \ln(x+1)\no\\
&-A_0A_1\ [\ \h\zeta_2+\Li_2(x)+\ln2 \ln(- x)\ ]
+A_1A_0\ [\ \h\zeta_2+\Li_2(x)+\ln(-x)\ln(1-x)\  ]\no\\
&+A_0A_2\ [\ \zeta_2-\Li_2(-x)\ ]-
A_2A_0\ [\ \zeta_2-\Li_2(-x)-\ln (-x)\ \ln(1+x)\ ]\no\\
&+A_1A_2\ [\ \Li_{1,1}(x,-1)-\h(\ln^22-\z_2)\ ]\no\\
&+A_2A_1\ [\ \Li_{1,1}(-x,-1)-\ln2\ln(x+1)+\h(\ln^22-\z_2)\ ]\no\\
&+\h\ \ln^2(-x)\ A_0^2 +\h\ [\ln(1-x)-\ln2]^2\ A_1^2+\h\ \ln^2(x+1)\ A_2^2
+\ldots\ .\label{PHI1m}
\end{align}
The auxiliary variable $\delta_x$ appears when the integration path contains a point at which the integrand is not analytic. Then $\delta_x$ denotes the branch above and below the real axis: $\delta_x=+1$ for $x\in {\bf H}_+$ and $\delta_x=-1$ for $x\in {\bf H}_-$ \cite{Panzer}.

One can compare the two solutions $\Phi_0$ and $\Phi_i$ referring to the  two points $x=x_0$ and $x=x_i,\ i=1,2$, respectively.
The quotient of any two such solutions is a constant non--commutative 
series known as regularized zeta series  giving rise to an associator \cite{BrownFuchs}. 
By analytic continuation the connection matrix between the solution $\Phi_0$ and $\Phi_1$ is independent of $x$ and gives rise to the first (Drinfeld) associator:
\be\label{Drinfeld1}
Z^{(+1)}(A_0,A_1,A_2) =\Phi_1(x)^{-1}\ \Phi_0(x)\ .
\ee
Furthermore, the connection matrix between the solution $\Phi_0$ and $\Phi_2$ is independent of $x$ and gives rise to the second (Drinfeld) associator:
\be\label{Drinfeld2}
Z^{(-1)}(A_0,A_1,A_2) =\Phi_2(x)^{-1}\ \Phi_0(x)\ .
\ee
The two associators \req{Drinfeld1} and \req{Drinfeld2} are independent and will be constructed below. 

The computation of the regularized zeta series 
\be\label{DB}
Z^{(x_k)}(A_0,A_1,\ldots,A_l)=\sum_{w\in A^\ast}\zeta^{x_k}(w)\ w\ \ \ ,\ \ \ k=1,\ldots,l
\ee
which for $l=2$ and  $x_1=1$ and $x_2=-1$ gives rise  to \req{Drinfeld1} and \req{Drinfeld2}, respectively has been proposed in \cite{BrownFuchs} by defining the map $\zeta^{x_k}$:
\be\label{mapzeta}
w\mapsto \zeta^{x_k}(w)\ .
\ee
The latter is a homomorphism for the shuffle product, i.e.:
\be\label{Shuff}
\zeta^{x_k}(w\shuffle \tilde w)=\zeta^{x_k}(w \tilde w)+\zeta^{x_k}(\tilde w w)=
\zeta^{x_k}(w)\  \zeta^{x_k}(\tilde w)\ .
\ee
The map \req{mapzeta} has the following properties:
\bea\label{Rezept}
\zeta^{x_k}(w_k)&=&-\ln(x_0-x_k)\ ,\\[2mm]
\zeta^{x_k}(w_0)&=&\ln(x_k-x_0)\ ,\\[2mm]
\zeta^{x_k}(w)&=&L_w(x_k)\ ,\ w\notin A^\ast w_0\cup w_k A^\ast\ ,
\eea
with $w$ neither beginning in $w_k$  nor ending in $w_0$.
Together with the shuffle product \req{Shuff} the relations  \req{Rezept} are enough to compute the map $\zeta^{x_k}$ for any word $w\in A^\ast$
because every word $w$ can be written uniquely as a linear combination of shuffle products of the words $w_0$ and $w_k$ and words which neither begin in $w_k$  nor end in $w_0$.
From \req{Rezept} the lowest orders of \req{DB} up to words of length one can be determined immediately \cite{BrownFuchs}
\be\label{DBB}
Z^{(x_k)}(A_0,\ldots,A_l)=1+\ln(x_k-x_0)\ A_0+\sum_{1\leq i\neq k}
\ln\lf(\fc{x_k-x_i}{x_0-x_i}\ri)\ A_i-\ln(x_0-x_k)\ A_k+\ldots\ .
\ee

In the following by using \req{Rezept} for $l=2$ we compute the higher orders of \req{DBB}
containing words up to length three and find
\begin{align}
Z^{(+1)}(A_0,A_1,A_2)&=1-i\pi\ A_1+\ln2\ A_2-\fc{\pi^2}{2}\ A_1^2+\h\ \ln^22\ A_2^2\no\\
&-\zeta_2\ [A_0,A_1]+\h\zeta_2\ [A_0,A_2]-\h(\ln^22-\zeta_2)\ [A_1,A_2]\no\\
&-i\pi\ln2\ A_1A_2+\fc{i\pi^3}{6}\ A_1^3+\fc{1}{6}\ln^32\ A_2^3\no\\
&+2\zeta_3\ A_0A_1A_0-\fc{3}{2}\z_3\ A_0A_2A_0+\lf(\fc{5}{8}\zeta_3-\zeta_2\ln2\ri)\ A_2A_0A_1\no\\
&-\lf(\fc{13}{8}\zeta_3-\fc{3}{2}\zeta_2\ln2-\fc{i\pi^3}{12}\ri)\ A_1A_2A_0-
\lf(\fc{13}{8}\zeta_3-\fc{3}{2}\zeta_2\ln2\ri)\ A_0A_2A_1\no\\
&+\lf(\zeta_3-\fc{1}{2}\zeta_2\ln2\ri)\ A_2A_1A_0+
\lf(\zeta_3-\fc{3}{2}\zeta_2\ln2\ri)\ A_0A_1A_2+\z_3\ A_0A_1^2+\fc{1}{8}\z_3\ A_0A_2^2\no\\
&+\lf(\fc{5}{8}\zeta_3-\fc{i\pi^3}{12}\ri)\ A_1A_0A_2+\fc{3}{4}\z_3\ (A_2A_0^2+A_0^2A_2)-
\z_3\ (A_1A_0^2+A_0^2A_1)\no\\
&+\lf(\fc{7}{8}\z_3-\h\z_2\ln2+\fc{1}{6}\ln^32\ri)\ A_2A_1^2-\lf(\fc{1}{8}\z_3-\fc{1}{6}\ln^32\ri)\ A_2^2A_1\no\\
&+\lf(\fc{1}{8}\z_3-\fc{1}{2}\z_2\ln2\ri)\ A_2^2A_0+\lf(\z_3-\fc{i\pi^3}{6}\ri)\ A_1^2A_0-\lf(2\z_3-\fc{i\pi^3}{6}\ri)\ A_1A_0A_1\no\\
&-\lf(\fc{1}{4}\z_3-\h\z_2\ln2\ri)\ A_2A_0A_2+\lf(\fc{1}{4}\z_3-\h\z_2\ln2+\fc{1}{6}\ln^32\ri)A_2A_1A_2\no\\
&+\lf(\fc{7}{8}\z_3-\fc{7}{2}\z_2\ln2+\fc{1}{6}\ln^32+\h i\pi(\ln^22-\z_2)\ri)\ A_1^2A_2\no\\
&-\lf(\fc{7}{4}\z_3-\z_2\ln2+\fc{1}{3}\ln^32+\h i\pi(\ln^22-\z_2)\ri)\ A_1A_2A_1\no\\
&-\lf(\fc{1}{8}\z_3-\fc{1}{2}\z_2\ln2+\fc{1}{3}\ln^32+\h i\pi\ln^22\ri)\ A_1A_2^2+\ldots\ ,\label{Drin1}
\end{align}
and:
\begin{align}
Z^{(-1)}(A_0,A_1,A_2)&=1+i\pi\ A_0+\ln2\ A_1-\fc{\pi^2}{2}\ A_0^2+\h\ \ln^22\ A_1^2\no\\
&-\zeta_2\ [A_0,A_2]+\h\zeta_2\ [A_0,A_1]+\h(\ln^22-\zeta_2)\ [A_1,A_2]\no\\
&+i\pi\ln2\ A_1A_0-\fc{i\pi^3}{6}\ A_0^3+\fc{1}{6}\ln^32\ A_1^3\no\\
&-\lf(\fc{3}{2}\zeta_3-\fc{i\pi^3}{12}\ri)\ A_0A_1A_0
+\lf(2\zeta_3-\fc{i\pi^3}{6}\ri)\ A_0A_2A_0+\lf(\fc{5}{8}\zeta_3-\zeta_2\ln2\ri)\  A_1A_0A_2\no\\
&+\lf(\zeta_3-\fc{3}{2}\zeta_2\ln2\ri)\ A_0A_2A_1+
\lf(\zeta_3-\fc{1}{2}\zeta_2\ln2+\fc{i\pi}{2}(\ln^22-\zeta_2)\ri)\ A_1A_2A_0\no\\
&-\lf(\fc{13}{8}\zeta_3-\fc{3}{2}\zeta_2\ln2\ri)\ A_0A_1A_2-
\lf(\fc{13}{8}\zeta_3-\fc{3}{2}\zeta_2\ln2+\fc{i\pi}{2}(\ln^22-\zeta_2)\ri)\ A_2A_1A_0\no\\
&+\fc{5}{8}\zeta_3\ A_2A_0A_1+\fc{3}{4}\z_3\ A_0^2A_1+\fc{1}{8}\z_3\ A_0A_1^2+\z_3\ (A_0A_2^2-A_0^2A_2+A_2^2A_0)\no\\
&+\lf(\fc{7}{8}\z_3-\h\z_2\ln2+\fc{1}{6}\ln^32\ri)\ A_1A_2^2-\lf(\fc{1}{8}\z_3-\fc{1}{6}\ln^32\ri)\ A_1^2A_2\no\\
&+\lf(\fc{1}{8}\z_3-\fc{1}{2}\z_2\ln2+\h i\pi\ln^22\ri)\ A_1^2A_0-2\z_3\ A_2A_0A_2\no\\
&-\lf(\fc{1}{4}\z_3-\h\z_2\ln2\ri)\ A_1A_0A_1 +\lf(\fc{1}{4}\z_3-\h\z_2\ln2+\fc{1}{6}\ln^32\ri)A_1A_2A_1\no\\
&+\lf(\fc{3}{4}\z_3-3\z_2\ln2-\h i\pi\z_2\ri)\ A_1A_0^2-\lf(\z_3-\fc{i\pi^3}{6}\ri)\ A_2A_0^2\no\\
&+\lf(\fc{7}{8}\z_3-\h\z_2\ln2+\fc{1}{6}\ln^32\ri)\ (A_2^2A_1-2A_2A_1A_2)\no\\
&-\lf(\fc{1}{8}\z_3-\h\z_2\ln2+\fc{1}{3}\ln^32\ri)\ A_2A_1^2+\ldots\ .\label{Drin2}
\end{align}

Since $\Phi_1(1)_{reg.}=1$ the associator $Z^{(+1)}$ can be related to the 
regularized value\footnote{The regularization has to be elaborated by properly taking into account the branches of the logarithm. E.g. at weight three we need: $\lf.\ln(1-x)\ri|_{x=1}=\lf.\ln(x-1)\ri|_{x=1}-i\pi\equiv-i\pi$, $\Li_{1,1}(1,-1)_{reg.}=\h\z_2-\h\ln^22-i\pi\ln2$, $\Li_{1,2}(1,1)_{reg.}=-2\z_3+\fc{i\pi^3}{6}$, $\Li_{1,2}(1,-1)=\fc{5}{8}\z_3-\fc{i\pi^3}{12}$, $\Li_{1,1,1}(1,1,-1)=-\fc{7}{8}\z_3+\fc{7}{2}\z_2\ln2-\fc{1}{6}\ln^32-\h i\pi(\ln^22-\z_2)$, $\Li_{1,1,1}(1,-1,-1)=\fc{7}{4}\z_3-\z_2\ln2+\fc{1}{3}\ln^32+\h i\pi(\ln^22-\z_2)$ and $\Li_{1,1,1}(1,-1,1)=\fc{1}{8}\z_3-\fc{1}{2}\z_2\ln2+\fc{1}{3}\ln^32+\h i\pi\ln^22$.}   of $\Phi_0$ at $x=1$, i.e.
\be\label{drinfeld1}
Z^{(+1)}=\Phi_0(1)_{reg.}\ ,
\ee
On the other hand, since $\Phi_2(-1)_{reg.}=1$ the associator $Z^{(-1)}$ can  be connected to the regularized value of $\Phi_0$ at $x=-1$, i.e.:
\be\label{drinfeld2}
Z^{(-1)}=\Phi_0(-1)_{reg.}\ .
\ee
Eventually, by combining Eq. \req{Drinfeld1}  with \req{drinfeld2} we find:
\be\label{Nice}
\Phi_1(-1)_{reg.}=Z^{(-1)}\ \ (Z^{(+1)})^{-1}\ .
\ee
Besides from $\Phi_0(0)_{reg.}=1$ we get:
\be
\Phi_1(0)_{reg.}=(Z^{(+1)})^{-1}\ \ \ ,\ \ \ \Phi_2(0)_{reg.}=(Z^{(-1)})^{-1}\ .
\ee
Furthermore, we also find
\be\label{Nicee}
\Phi_2(1)_{reg.}=Z^{(+1)}\ \ (Z^{(-1)})^{-1}\ ,
\ee
which in turn implies:
\be
\Phi_1(-1)_{reg.}=\Phi_2(1)_{reg.}^{-1}\ .
\ee
Finally, with the explicit expressions \req{Drin1} and \req{Drin2} we compute
the product entering \req{Nice}:
\begin{align}
Z^{(-1)}\ (Z^{(+1)})^{-1}&=1+\ln2\ (A_1-A_2)+(i\pi)\ (A_0+A_1)
+\h(i\pi)^2\ (A_0^2+A_1^2+2A_0A_1)\no\\
&+\h\ln^22\ (A_1^2+A_2^2-2A_1A_2)
+\fc{3}{2}\zeta_2\ [A_0,A_1]-\fc{3}{2}\zeta_2\ [A_0,A_2]+(\ln^22-\zeta_2)\ [A_1,A_2]
\no\\
&+(i\pi)\ln2\ \ (A_1A_0-A_0A_2-A_2A_1)+\fc{1}{6}(i\pi)^3\ A_0^3+\fc{1}{6}\ln^3(-2)\ A_1^3-\fc{1}{6}\ln^32\ A_2^3\no\\
&-\lf(\fc{7}{2}\zeta_3+\fc{i\pi^3}{12}\ri)\ A_0A_1A_0+\lf(\fc{7}{2}\zeta_3-\fc{i\pi^3}{12}\ri)\ A_0A_2A_0-(i\pi)\ \ln^22\ A_1A_0A_2\no\\
&+\lf(\fc{21}{8}\zeta_3+3\zeta_2\ln2-(i\pi)\ (\zeta_2+\fc{1}{2}\ln^22)\ri)\ A_0A_2A_1\no\\
&+\lf(\fc{21}{8}\zeta_3-\fc{3}{2}\zeta_2\ln2+\h(i\pi)\ (\ln^22-\zeta_2) \ri)\ A_1A_2A_0\no\\
&+\lf(-\fc{21}{8}\zeta_3+\fc{3}{2}\zeta_2\ln2+\h(i\pi)\ (\ln^22-\zeta_2) \ri)\ A_0A_1A_2\no\\
&+\lf(-\fc{21}{8}\zeta_3+3\zeta_2\ln2-\h(i\pi)\ (\ln^22-\zeta_2) \ri)\ A_2A_1A_0
+\fc{3}{2}(i\pi)\ \zeta_2\ A_2A_0A_1\no\\
&+\lf(\fc{7}{4}\z_3-\fc{i\pi^3}{3}\ri)\ A_0^2A_1-\lf(\fc{7}{4}\z_3+\fc{i\pi^3}{12}-3\z_2\ln2\ri)\ A_0^2A_2\no\\
&+\lf(\fc{7}{8}\z_3+\fc{i\pi}{2}\ln^22+\fc{3}{2}\z_2\ln2\ri)\ A_0A_2^2+\lf(-\fc{7}{8}\z_3-\fc{i\pi^3}{4}\ri)\ A_0A_1^2+\z_3\ (A_1A_2^2-A_1^2A_2)\no\\
&+\lf(2\z_3-\z_2\ln2-\fc{i\pi^3}{6}\ri)\ A_1A_2A_1+\lf(\fc{7}{4}\z_3+\fc{3}{2}\z_2\ln2-6\z_2\ln2-\fc{i\pi^3}{4}\ri)\ A_1A_0A_1\no\\
&+\lf(\fc{7}{4}\z_3-3\z_2\ln2-\fc{i\pi^3}{12}\ri)\ A_1A_0^2-\lf(\fc{7}{4}\z_3-\fc{i\pi^3}{6}\ri)\ A_2A_0^2\no\\
&-\lf(\fc{7}{8}\z_3+\fc{3}{2}\z_2\ln2-\h(i\pi)\ln^22\ri)\ A_1^2A_0+\fc{7}{8}\z_3\ A_2^2A_0-
\lf(\fc{7}{4}\z_3+\fc{3}{2}\z_2\ln2\ri)\ A_2A_0A_2\no\\
&+\lf(\z_3-\z_2\ln 2+\h\ln^32+\h(i\pi)\ln^22\ri)\ A_2^2A_1-\lf(2\z_3-\z_2\ln2\ri)\ A_2A_1A_2\no\\
&-\lf(\z_3-4\z_2\ln2+\h\ln^32+(i\pi)(\ln^22-\z_2)\ri)\ A_2A_1^2+\ldots\ .\label{NICE}
\end{align}

\subsection{Hypergeometric function expansion from  associators}
\def\F{{}_2F_1}

In this subsection we relate our generic solutions to \req{Schlesinger} to 
$\eps$--expansions of hypergeometric functions with half--integer parameters.

For $a,b,c+\h-a, c+\h-b$ and $c+\h$ non--integers a pair of two independent solutions to \req{3hgdgl} is characterized by their local exponents at the regular singularities $0,1$ and $\infty$ 
 described by the Riemann scheme
\be\label{scheme}
\begin{pmatrix}
\underline{0} &\underline{1} &\underline{\infty}   \\  
0&0&a\\
\h-c&c+\h-a-b&b
\end{pmatrix}
\ee
E.g. at $z=0$ we can choose the two independent solutions:
\bea
u_1&=&\ds{\F\lf[{a,b\atop c+\h};z\ri]}\\[5mm]
u_2&=&\ds{z^{\h-c}\ \F\lf[{a+\h-c\ ,\ b+\h-c\atop \fc{3}{2}-c};z\ri]\ ,}\label{gMatrix}
\eea
which can be combined into the matrix
\be\label{solmatrix}
\bm{g}=\begin{pmatrix}
u_1&u_2\\
\th u_1&\th u_2
\end{pmatrix}\ ,
\ee
which for $z\ra0$ behaves as:
\be\label{limit1}
\bm{g}\stackrel{z \ra 0}{\lra} \begin{pmatrix}
1&z^{\h-c}\\[5mm]
0&\lf(\h-c\ri)\ z^{\h-c}\\
\end{pmatrix}\ .
\ee
The set of solutions \req{gMatrix} can be related to the normalized solution $\Phi_1(y)$ given in \req{PHI1} subject to the transformation \req{subst}, i.e.:
\be\label{Subst}
z=-\fc{(1-y)^2}{4y}\ .
\ee
This is accomplished by introducing the connection matrix C:
\be\label{connection}
\bm{g}=\Phi_1\ C\ .
\ee
By construction for $y\ra1$ the solution $\Phi_1(y)$  behaves as:
\be\label{limit2}
\Phi_1(y)\lra (y-1)^{A_1}=\begin{pmatrix}
1&0\\[5mm]
0&(y-1)^{-2c}
\end{pmatrix}\ .
\ee
Comparing \req{limit1} and \req{limit2} gives the following connection matrix:
\be\label{Cmatrix}
C=\begin{pmatrix}
1&0\\[5mm]
0&(y-1)^{2c}
\end{pmatrix}\ 
\begin{pmatrix}
\ds{1}&\ds{\h\ (-1)^{\h-c} 4^c\ (1-y)^{1-2c} }\\[5mm]
\ds{0}&\ds{\h\ (-1)^{\h-c} 4^c\  (1-y)^{1-2c}\lf(\h-c\ri)}\\
\end{pmatrix}
=\begin{pmatrix}
1&\ast\\[5mm]
0&\ast
\end{pmatrix}\ .
\ee
The first column of $C$ does not depend on $z$ nor $y$, while the non--vanishing 
second column is not relevant to us.
In fact, with \req{connection} the form \req{Cmatrix} of $C$ allows us to relate $u_1$ of \req{gMatrix} to the first matrix element of the fundamental solution \req{PHI1} as
\be\label{FINAL1}
\F\lf[{a,b\atop c+\h};z\ri]=\lf.\Phi_1[A_0,A_1,A_2](y)\ \ri|_{1,1}\ ,
\ee
with the matrices $A_i$ given in \req{matrices} and subject to the relations \req{subst} and \req{Subst}.
Up to words of length four \req{FINAL1} gives
\begin{align}
\F\lf[{a,b\atop c+\h};z\ri]&=1+\h\  \ln^2y\ A_0^2 +\fc{1}{6}\ \ln^3y\ A_0^3\no\\
&+\lf[\ 2 \Li_3(y)-\ln y\Li_2(y)-\z_2 \ln y-2\z_3\ \ri]\  A_0A_1A_0\no\\
&+\lf[\ 2 \Li_3(-y)-\ln y\Li_2(-y)+\h \z_2\ \ln y+\fc{3}{2} \z_3\ \ri]\  A_0A_2A_0+\fc{1}{4!}\ \ln^4y\ A_0^4\no\\
&+\lf(\fc{6}{5} \z_2^2-3\ \Li_4(y)-\h \ln^2y \Li_2(y)+2 \ln y\Li_3(y)+\z_3\ln y\ri)\ A_0A_1A_0^2\no\\
&-\lf(\fc{21}{20} \z_2^2+3\ \Li_4(-y)+\h \ln^2y \Li_2(-y)-2 \ln y\Li_3(-y)+\fc{3}{4} \z_3\ln y\ri)\ A_0A_2A_0^2\no\\
&-\lf(\fc{6}{5} \z_2^2-3 \Li_4(y)+\ln y\Li_3(y)+\h\z_2\ln^2y+2\z_3\ln y\ri)\ A_0^2A_1A_0\no\\
&-\lf(\h \z_2^2+\z_3\ln y-\z_2\Li_2(y)+\Li_{2,2}(y,1)+2\Li_{3,1}(y,1)-\ln y\Li_{2,1}(y,1)\ri)\ A_0A_1^2A_0\no\\
&-\lf(\fc{27}{40}\ \z_2^2+2\Li_{3,1}(-1,1)-\fc{7}{2}\z_3\ln2+\Li_{2,2}(-y,-1)+2\ \Li_{3,1}(-y,-1)\ri.\no\\
&-\lf.\ln y\Li_{2,1}(-y,-1)-\z_2\Li_2(-y)+\z_3\ln y-\fc{3}{2}\z_2\ln2\ln y\ri)\ A_0A_2A_1A_0\no\\
&+\lf(\fc{21}{20} \z_2^2+3 \Li_4(-y)-\ln y\Li_3(-y)+\fc{1}{4}\z_2\ln^2y+\fc{3}{2}\z_3\ln y\ri)\ A_0^2A_2A_0\no\\
&+\lf(\fc{47}{40} \z_2^2+2\Li_{3,1}(-1,1)-\fc{7}{2}\z_3\ln2-\Li_{2,2}(y,-1)-2\ \Li_{3,1}(y,-1)\ri.\no\\
&+\lf.\ln y\Li_{2,1}(y,-1)-\h\z_2\Li_2(y)+\fc{13}{8}\z_3\ln y-\fc{3}{2}\z_2\ln2\ln y\ri)\ A_0A_1A_2A_0\no\\
&-\lf(\fc{1}{8} \z_2^2+\h\z_2\Li_2(-y)+\Li_{2,2}(-y,1)-\ln y\Li_{2,1}(-y,1)+2\Li_{3,1}(-y,1)\ri.\no\\
&+\lf.\lf.\fc{1}{8}\z_3\ln y\ri)\ A_0A_2^2A_0\ \ri|_{1,1}+\ldots\ ,\label{fINAL1}
\end{align}
which exactly agrees with the orders displayed in \req{1stord}.

Finally, with \req{Nice} we can express the function \req{Start} evaluated at $z=1$ 
in terms of the two independent associators:
\be\label{FINAL2}
\F\lf[{a,b\atop c+\h};1\ri]=\lf.Z^{(-1)}\ \ (Z^{(+1)})^{-1}\ \ri|_{1,1}\ .
\ee
The r.h.s. of \req{FINAL2} can be found in \req{NICE}.
Up to words of length four \req{FINAL2} yields
\begin{align}
\F\lf[{a,b\atop c+\h};1\ri]&=1+\h\ (i\pi)^2\ A_0^2+\fc{1}{6}\ (i\pi)^3\ A_0^3\no\\
&-\lf(\fc{7}{2}\zeta_3+\fc{i\pi^3}{12}\ri)\ A_0A_1A_0+\lf(\fc{7}{2}\zeta_3-\fc{i\pi^3}{12}\ri)\ A_0A_2A_0 \no\\
&+\fc{3}{2}\z_2^2\ A_0^4+\lf(\fc{3}{4}\z_2^2-\h (i\pi)\z_3\ri)\ A_0A_1A_0^2+
\lf(\fc{3}{4}\z_2^2+\fc{5}{4} (i\pi)\z_3\ri)\ A_0A_2A_0^2\no\\
&+\lf(\fc{3}{4}\z_2^2-\fc{5}{4} (i\pi)\z_3\ri)\ A_0^2A_1A_0-
\lf(\fc{9}{8}\z_2^2+\fc{7}{8} (i\pi)\z_3\ri)\ A_0A_1^2A_0\no\\
&-\lf(\fc{7}{20}\z_2^2+4\Li_{3,1}(-1,1)-7\zeta_3\ln 2\ri)\ A_0A_2A_1A_0+
\lf(\fc{3}{4}\z_2^2+\h (i\pi)\z_3\ri)\ A_0^2A_2A_0\no\\
&\lf.+\lf(\fc{13}{5}\z_2^2+4\Li_{3,1}(-1,1)-7\zeta_3\ln 2\ri)\ A_0A_1A_2A_0-
\lf(\fc{9}{8}\z_2^2-\fc{7}{8}(i\pi)\z_3\ri)\ A_0A_2^2A_0+\ldots\ri|_{1,1}\no\\
&=1+\h\ \zeta_2\ ab+7\ \z_3\ ab\ (a+b-2c)\no\\
&+\fc{3}{2}\ \z_2^2\ ab\ (4a^2+9ab+4b^2-12ac-12bc+12c^2)+\ldots\ ,\label{fINAL2}
\end{align}
where we have used
\be
\begin{array}{rllll}
\lf.A_0^2\ri.|_{1,1}&=-ab\ ,&\ \lf.A_0^3\ri.|_{1,1}&=-ab\ (a+b)\ ,\\[3mm]
\lf.A_0A_1A_0\ri.|_{1,1}&=2\ abc\ ,&\ \lf.A_0A_2A_0\ri.|_{1,1}&=2\ ab\ (a+b-c)\ ,
\end{array}
\ee
and the non--vanishing fourth order contributions:
\be
\begin{array}{rllll}
\lf.A_0^4\ri.|_{1,1}&=-ab\ (a^2+ab+b^2)\ ,&\ \lf.A_0A_1A_0^2\ri.|_{1,1}&=2abc\ (a+b)\ ,\\[3mm]
\lf.A_0A_2A_0^2\ri.|_{1,1}&=2ab\ (a+b)\ (a+b-c)\ ,&\ \lf.A_0^2A_1A_0\ri.|_{1,1}&=2abc\ (a+b)\ ,\\[3mm]
\lf.A_0A_1^2A_0\ri.|_{1,1}&=-4abc^2\ ,&\ \lf.A_0A_2A_1A_0\ri.|_{1,1}&=-4abc\ (a+b-c)\ ,\\[3mm]
\lf.A_0^2A_2A_0\ri.|_{1,1}&=2ab\ (a+b)\ (a+b-c)\ ,&\ \lf.A_0A_1A_2A_0\ri.|_{1,1}&=-4abc\ (a+b-c)\ ,\\[3mm]
\lf.A_0A_2^2A_0\ri.|_{1,1}&=-4ab\ (a+b-c)^2\ .&
\end{array}
\ee
It is worth pointing out the compactness and simplicity of the results 
\req{fINAL1} and \req{fINAL2}. For a given 
weight $k$ in \req{3hg} the coefficients $u_k(z)$ are given by the MPLs of degree $k$ entering the fundamental solution \req{PHI0} (subject to \req{subst}) supplemented 
by matrix products of $k$ matrices $A_0,A_1$ and $A_2$ given in \req{matrices}.
Furthermore, for a given degree $k$ the coefficients $u_k(1)$ are given by the terms of degree $k$ of 
product \req{NICE} of the two universal associators \req{Drin1} and \req{Drin2}.
As a consequence, computing higher orders in the expansion \req{3hg} is reduced to simple matrix multiplications.


\section{A simple Feynman integral}

We use the family of one--loop bubble integrals
\begin{align}\label{fam}
I(a_1,a_2)=\int\frac{d^Dk}{i\pi^{D/2}}\frac{1}{D_1^{a_1}D_2^{a_2}} \ , \ \ a_1,a_2\in \mathbb{Z} \ ,
\end{align}
with massive propagators
\begin{align*}
D_1&=k^2-m^2+i0\ ,\\
D_2&=(k+p)^2-m^2+i0\ ,
\end{align*}
and $p^2=s$ as a simple example to demonstrate, how to obtain all--order expansions in $\epsilon$ ($D=4-2\epsilon$) for Feynman integrals. This is achieved for a MI of the family \req{fam} in section 3.1. In section 3.2 we discuss how to obtain the representation for that MI in terms of a hypergeometric function from the all--order expansion.

The integral $I(a_1,a_2)$ is well--known. It is given in \cite{Davy} for general dimensions in terms of generalized hypergeometric functions. Furthermore, various representations of $I(1,1)$ are discussed in \cite{Davydychev:2000na}. This includes all--order expansions, for example in terms of Nielsen polylogarithms, and representations in terms of hypergeometric functions ${}_2F_1$. Both references also discuss the case with propagators of different masses.

\subsection[All--order Laurent expansion in $\epsilon$]{All--order Laurent expansion in $\bm{\epsilon}$}

\begin{figure}[H]
\centering
\includegraphics[width=0.5\textwidth]{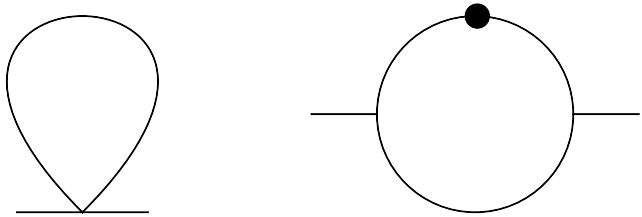}
\caption{Master integrals of topology \req{fam}: The massive tadpole integral (left) $I(1,0)$ and the massive bubble integral (right) with the dot representing a squared propagator.}
\label{MIs}
\end{figure}

IBP identities \cite{Smirnov:2008iw} allow to express any integral $I(a_1,a_2)$ in terms of two MIs (cf. Fig. \ref{MIs}).

\noindent
A natural choice for one of the MIs is the massive tadpole integral $I(1,0)$, which follows from the well--known result
\begin{align}\label{tad}
I(a,0)=(-1)^{a}\frac{\Gamma(a+\epsilon-2)}{\Gamma(a)}(m^2)^{2-\epsilon-a}
\end{align}
 for $a=1$. As second MI we consider the finite integral $I(2,1)$. We use the Ansatz
\begin{align}\label{ans}
I(2,1)=\Gamma(\epsilon+1)\frac{(m^2)^{-\epsilon}}{\sqrt{s(s-4m^2)}}\; g(x)\ ,
\end{align}
where the function $g(x)$ depends on the ratio of the squared mass $m^2$ and the kinematic invariant $s$ through the variable:
\begin{align}\label{x}
x=\frac{\sqrt{1-\frac{4m^2}{s}}-1}{\sqrt{1-\frac{4m^2}{s}}+1}\ .
\end{align}
In the region $s<0$, which transforms to $0<x<1$ under eq. \req{x}, the function $g(x)$ satisfies the following first order differential equation in $x$ \cite{Kotikov:1991pm}:
\begin{align}\label{diff1}
\frac{d}{dx}g(x)=\left(\frac{1}{x}-\frac{2}{1+x}\right)\epsilon\; g(x)+\frac{1}{x} \ .
\end{align}
Instead of solving this equation directly we consider the Laurent expansion around $\epsilon=0$:
\begin{align}\label{exp}
g(x)=\sum\limits_{k\geq0}g_k(x)\ \epsilon^k \ .
\end{align}
Inserting \req{exp} into eq. \req{diff1} and using the fact, that the resulting differential equation is valid at any order in $\epsilon$, yields first order differential equations for the coefficient functions $g_k(x)$:
\begin{align}
\begin{split}
\frac{d}{dx} g_0(x)&=\frac{1}{x}\ ,\\
\frac{d}{dx} g_k(x)&=\left(\frac{1}{x}-\frac{2}{1+x}\right)g_{k-1}(x)\ , \  \ k\geq 1\ .\label{diff}
\end{split}
\end{align}
We use the limit of vanishing external momenta, i.e. $s=0$ (which corresponds to $x=1$), in $I(2,1)\stackrel{s= 0}{\rightarrow} I(3,0)$ and eq. \req{tad} to determine the boundary conditions $g_k(1)=0$. This allows to straightforwardly solve the differential equations \req{diff}. We can directly calculate $g_0(x)$ and give an iterative solution for all other coefficient functions:
\begin{align}
g_0(x)&=H(0;x)\label{o0}\ ,\\
g_k(x)&=\int_{1}^x\frac{dx'}{x'}g_{k-1}(x')-2\int_1^x\frac{dx'}{1+x'}g_{k-1}(x')\ ,\ \ k\geq 1\ .\label{it}
\end{align}
For example  we obtain the next orders in terms of HPLs:
\begin{align}\label{iv}
g_1(x)&=H(0,0;x)-2H(-1,0;x)-\zeta(2)\ ,\\
g_2(x)&=H(0,0,0;x)-2H(0,-1,0;x)-2H(-1,0,0;x)+4H(-1,-1,0;x)\nonumber\\
&-\zeta(2)H(0;x)+2\zeta(2)H(-1;x)-2\zeta(3)\ .\label{o2}
\end{align}
Representing the Feynman integral by the lowest orders of the $\epsilon$--expansion is sufficient to describe the behaviour around $\epsilon=0$. But not all informations on the integral are contained in a finite number of orders. Therefore we are interested in the all--order expression for $g_k(x)$: an equation valid for all $k$, which in contrast to eq. \req{it} is given explicitly in terms of iterated integrals. An expression for $g_k(x)$ of this type would allow to give the $\epsilon$--expansion \req{exp} as an infinite series thereby providing a closed representation of the complete Feynman integral.

Applying the definitions of the integral operators \req{3jops} in the iterative solution \req{it} yields the first order recurrence relation:
\begin{align}\label{frec}
g_k(x)=\left[J(0)-2J(-1)\right]g_{k-1}(x)\ , \ \ k\geq1  \ .
\end{align}
With the general solution for this type of relations given in \cite{StSt} and the initial value \req{o0} the all--order expression
\begin{align}\label{res}
g_k(x)=\sum\limits_{j_1+j_2=k}(-2)^{j_2}\{J(0)^{j_1},J(-1)^{j_2}\}J(0)
\end{align}
follows directly. The sum is over non--negative integers $j_1,j_2$ which fulfil the condition $j_1+j_2=k$. 
While eq. \req{it} depends on lower orders of the expansion, eq. \req{res} allows to extract the coefficient function of any order in $\eps$ directly. For example with $k=2$ the condition in the sum over $j_1,j_2$ has three solutions, which give the following terms:
\begin{align}
\begin{split}
(j_1,j_2)=&(2,0):\ \ J(0,0,0)\ ,\\
&(1,1):\ \ -2J(0,-1,0)-2J(-1,0,0)\ ,\\
&(0,2):\ \ 4J(-1,-1,0)\ .
\end{split}
\end{align}
In agreement with eq. \req{o2} the sum of all three terms gives $g_2(x)$. 

We can now give a compact expression for $g(x)$ in terms of integral operators:
\begin{align}\label{fres}
g(x)=\sum\limits_{j_1,j_2\geq0}\epsilon^{j_1+j_2}(-2)^{j_2}\{J(0)^{j_1},J(-1)^{j_2}\}J(0)\ .
\end{align}
Using the recursive definition of the generalized operator product given in \cite{StSt}, it easy to prove, that \req{fres} solves the differential equation \req{diff1}. On the other hand, eq.  \req{it} or the equivalent recurrence relation \req{frec} only solve the differential eq. \req{diff} iteratively.

Alternatively to study a Laurent expansion in $\epsilon$ in terms of iterated integrals one can look for a solution exact in $\epsilon$ in terms of hypergeometric functions and their generalizations. However, for the latter the behaviour of the integral at $\epsilon=0$ is not obvious. The all--order representation combines the advantages of both approaches. It is given explicitly in terms of iterated integrals, which allows to straightforwardly determine any order of the $\epsilon$--expansion and therefore the behaviour at $\epsilon=0$. Moreover, the all--order result contains all informations on the Feynman integral in a compact form like a representation in terms of hypergeometric functions. In the next section we give an application for the all--order expression, which confirms this statement.

\subsection{Representation in terms of a hypergeometric function}

Starting with expression \req{fres} it is possible to construct the representation of $I(2,1)$ in terms of a hypergeometric function, because the all--order expression of the latter is known as well. The type of integral operators which are part of \req{fres} also appear in the expansion of the hypergeometric function ${}_2F_1$, which we have considered in section 2. Combining eqs. \req{3hg} and \req{3half} this expansion can be written as:
\begin{multline}\label{hgc}
{}_2F_1\left[{a\epsilon,b\epsilon\atop\frac{1}{2}+c\epsilon};z\right]=\sum\limits_{l_1,l_2,l_3,m\geq0}\epsilon^{l_1+l_2+l_3+2m+2}(-1)^{l_2+m+1}2^{l_1+l_2}c^{l_1}(a+b-c)^{l_2} (a+b)^{l_3}(ab)^{m+1}\\
\times J(0)\{J(1)^{l_1},J(-1)^{l_2},J(0)^{l_3},J(0,0)^m\}J(0)\ .
\end{multline}
Let us start by comparing the products of integral operators in the all--order expression \req{fres} and \req{hgc}. One difference is the operator $J(0)$ to the left of the generalized operator product in \req{hgc}, which is missing in \req{fres}. Therefore we shall deal  with a hypergeometric function, which is the product of the differential operator $y\frac{d}{dy}$ and \req{hgc}. According to eqs. \req{3hgrel} and \req{subst} this function can be written in the form:
\begin{align}
y\frac{d}{dy}\;{}_2F_1\left[{a\epsilon,b\epsilon\atop\frac{1}{2}+c\epsilon};z\right]=-\frac{1+y}{1-y}\; \frac{zab\epsilon^2}{\frac{1}{2}+c\epsilon}\; {}_2F_1\left[{1+a\epsilon,1+b\epsilon\atop\frac{3}{2}+c\epsilon};z\right]\ .
\end{align}
Inserting the expansion \req{hgc} on the l.h.s. we arrive at 
\begin{multline}\label{hgd}
{}_2F_1\left[{1+a\epsilon,1+b\epsilon\atop \frac32+c\epsilon};z\right]=\frac{4y}{1-y^2}\left(\frac12+c\epsilon\right)\sum\limits_{l_1,l_2,l_3,m\geq0}\epsilon^{l_1+l_2+l_3+2m}(-1)^{l_2+m+1}2^{l_1+l_2}\\ \times c^{l_1} (a+b-c)^{l_2}(a+b)^{l_3}(ab)^m\{J(1)^{l_1},J(-1)^{l_2},J(0)^{l_3},J(0,0)^m\}J(0)\ ,
\end{multline}
where identical to eq. \req{fres} the operator $J(0)$ to the left of the generalized operator product is now missing. In the next step we want to remove the operators $J(1)$ and $J(0,0)$ in \req{hgd}, which do not appear in \req{fres}. To achieve this, we set $a=c=0$ so that only the part of the sum with $l_1=m=0$ remains:
\begin{align}\label{hge}
{}_2F_1\left[{1,1+b\epsilon\atop \frac32};z\right]=-\frac{2y}{1-y^2}\sum\limits_{l_2,l_3\geq0}\epsilon^{l_2+l_3}(-2)^{l_2}b^{l_2+l_3}\{J(-1)^{l_2},J(0)^{l_3}\}J(0)\ .
\end{align}
Now the integral operators are identical to the ones in \req{fres}. To adjust the other factors, we set $b=1$ and replace $y$ and $z$ via
\begin{align}
y=x \ , \  \ z=\frac{s}{4m^2}\ ,
\end{align}
which yields the final result:
\begin{align}\label{fg}
g\left(\frac{s}{m^2}\right)=-\frac{\sqrt{s(s-4m^2)}}{2m^2}{}_2F_1\left[{1,1+\epsilon\atop\frac32};\frac{s}{4m^2}\right]\ .
\end{align}
From eq. \req{ans} follows the expression for the complete integral:
\begin{align}
I(2,1)=-\frac12\;\Gamma(\epsilon+1)(m^2)^{-\epsilon-1}{}_2F_1\left[{1,1+\epsilon\atop\frac32};\frac{s}{4m^2}\right]\ .
\end{align}
With this representation we can make use of the fact that we obtained an expression exact in $\epsilon$.  Replacing $\epsilon$ with $2-\frac{D}{2}$ yields the Feynman integral in general dimensions $D$:
\begin{align}
I_{D}(2,1)=-\frac12\;\Gamma\left(3-\frac{D}{2}\right)(m^2)^{\frac{D}{2}-3}{}_2F_1\left[{1,3-\frac{D}{2}\atop\frac32};\frac{s}{4m^2}\right]\ .
\end{align}
This result is identical to the one presented in Ref. \cite{Davy} for the region $s<4m^2$, cf. also 
\cite{Anastasiou:1999ui}.

\section{Concluding remarks}

We may extend our analysis  to  hypergeometric functions
${}_2F_{1}$ with parameters shifted by rational values $p/q$.
Then, in lines of \req{possa} we have the four cases
\begin{align}
&{}_2F_{1}\left[{a,b\atop 1-\fc{p}{q}+c};z\ri]\ ,\label{Possa}\\[5mm]
\ {}_2F_{1}\left[{1-\fc{p}{q}+a,1-\fc{p}{q}+b\atop 1-\fc{p}{q}+c};z\ri],\ & 
{}_2F_{1}\left[{1-\fc{p}{q}+a,b\atop 1-\fc{p}{q}+c};z\ri],\ {}_2F_{1}\left[{1-\fc{p}{q}+a,b\atop 1+c};z\ri]\ ,\label{Possaa}
\end{align}
which boil down to \req{possa} for $\fc{p}{q}=\h$. Again, additional integer shifts in their parameters can be treated in the same way as for the case 
$\fc{p}{q}=\h$ described in the beginning of section 2.
The all order $\eps$--expansions of
the first three functions  involve MPLs of $q$--th roots of unity:
\be\label{roots}
r_l=\exp\lf(\fc{2\pi i l}{q}\ri)\ \ \ ,\ \ \ l=1,\ldots,q\ .
\ee 
On the other hand, the expansion of the last function is expressible in terms of  MPLs of $q$--th roots of unity times powers of logarithms $\ln(1-z)$  \cite{Kalmykov:2008ge}.
Again, according to \cite{Kalmykov:2008ge} any of the three functions \req{Possaa} can be expressed in terms of the first one \req{Possa}. Therefore, it is sufficient to study 
the $\eps$--expansion of \req{Possa}. 

After introducing the new variable  \cite{Weinzierl:2004bn,Kalmykov:2008ge}
\be\label{VARN}
y=\lf(\fc{z}{z-1}\ri)^{1/q}\ ,
\ee
the differential equation \req{3hgdgl} 
\begin{align}\label{3hgdglnew}
z(\theta+a)(\theta+b)\;{}_2F_{1}\left[{a,b\atop c+1-\fc{p}{q}};z\right]=\theta\ (\theta+c-\fc{p}{q})\ {}_2F_{1}\left[{a,b\atop c+1-\fc{p}{q}};z\right]
\end{align}
for the function \req{Possa} 
\be
\varphi_1(y):={}_2F_{1}\left[{a,b \atop c+1-\fc{p}{q}};z \right]
\ee
can be written as a system of two first order differential equations \cite{Kalmykov:2008ge}:
\begin{align}
\fc{1}{q}\ \fc{d}{dy}\ \varphi_1(y) &=\fc{y^{p-1}}{1-y^q}\ \varphi_2(y)\ ,
\no \\
-\fc{1}{q}\ \fc{d}{dy} \varphi_2(y)&=\lf[\ (a_1+a_2)\ \fc{y^{q-1}}{1-y^q}+
\fc{c}{y}\ \ri]\ \varphi_2(y)+a_1a_2\ \fc{y^{q-p-1}}{1-y^q}\ \varphi_1(y)\ .\label{DGLN}
\end{align}
For $p<q$ the rational functions in \req{DGLN} can be decomposed by means of partial fraction decomposition as
\begin{align}
\fc{y^{q-1}}{1-y^q}&=-\fc{1}{q}\ \sum_{l=1}^q\fc{1}{y-r_l}\ ,\no\\
\fc{y^{p-1}}{1-y^q}&=-\fc{1}{q}\ \sum_{l=1}^q\fc{r_l^p}{y-r_l}\ ,\no\\
\fc{y^{q-p-1}}{1-y^q}&=-\fc{1}{q}\ \sum_{l=1}^q\fc{r_l^{-p}}{y-r_l}\ ,
\end{align}
with the $q$--th roots of unity \req{roots}.
Then  the system \req{DGLN} can be brought into Fuchsian class
\be\label{SCHlesinger}
\fc{d\Phi}{dy}=\lf(\sum_{i=0}^q\fc{A_i}{y-y_i}\ri)\ \Phi\ ,
\ee
with the vector
\be
\Phi=\lf({\varphi_1\atop \varphi_2}\ri)\ ,
\ee
and the $2\times2$ matrices $A_i$:
\be\label{Matrices}
A_0=\begin{pmatrix}
0&0\\
0&-q\ c
\end{pmatrix}\ \ \ ,\ \ \ A_l=\begin{pmatrix}
0&-r_l^p\\
ab\ r_l^{-p}&a+b
\end{pmatrix}\ \ \ ,\ \ \ l=1,\ldots,q\ .
\ee
The system \req{SCHlesinger} has $q+2$ singularities at $y_0=0,\ y_i=r_i,\ i=1,\ldots,q$ and $\infty$.   
Note, that for $q=2$ the matrices do not boil down to \req{matrices} because the transformation \req{VARN} differs from \req{subst}. Nevertheless, for this case the system \req{SCHlesinger} represents an equivalent representation, which becomes manifest in its solutions subject to the transformation rules  of the HPLs
 under $y\mapsto\fc{1-y}{1+y}$.

In the generic case $q>2$
under consideration  with the $q+1$ singular points $x_0=0$ and $x_i=r_i,\ i=1,\ldots,q$
the alphabet $A$ consists of $q+1$ letters $A^\ast=\{w_0,w_1,w_2,\ldots w_q\}$ and is directly related to the differential forms 
$$\fc{dx}{x}\ \ ,\ \ \bigcup_{l=1}^q\fc{dx}{x-r_l}$$ 
appearing in \req{SCHlesinger}. Now the corresponding hyperlogarithms \req{hyperlog} 
involve the $q$--th roots of unity \req{roots}. 
Again, a grouplike solution $\Phi$ to \req{SCHlesinger}  taking values in ${\bf C}\vev{A}$  with the alphabet 
$A=\{A_0,A_1,\ldots,A_q\}$ can be given as formal weighted sum over iterated integrals  (with the weight given by the number of iterated integrations)
\be\label{genSOLL}
\Phi(x)=\sum_{w\in A^\ast}L_w(x)\ w\ ,
\ee
with the MPLs \req{hyperlog}. 
The solution \req{genSOLL} can be constructed recursively and built by Picard's iterative methods.
Again, one can construct $q+1$ unique analytic solutions $\Phi_i$ normalized at $x=x_i$ with the asymptotic behaviour $x\ra x_i$ as
\be
\Phi_i(x)\lra (x-x_i)^{A_i}\ \ \ ,\ \ \ i=0,\ldots,q\ .
\ee
Their $q$ regularized zeta series \req{DB} corresponding to the ratios
\be
Z^{(x_i)}(A_0,A_1,\ldots,A_q)=\Phi_i(x)^{-1}\ \Phi_0(x)\ \ \ ,\ \ \ i=1,\ldots,q
\ee
can be computed along the lines presented above and their lowest orders are given in \req{DBB}.
Again, the latter can be related to \req{Possa} at special values
\cite{Progress}.

\vskip0.5cm
\goodbreak
\centerline{\noindent{\bf Acknowledgments} }\vskip 1mm
We wish to thank Stefan Weinzierl for an useful email exchange.


\begin{thebibliography}{10}

\bibitem{Weinzierl:2004bn}
  S.~Weinzierl,
``Expansion around half integer values, binomial sums and inverse binomial sums,''
  J.\ Math.\ Phys.\  {\bf 45} (2004) 2656
  [hep-ph/0402131].
 
   \bibitem{Slater} 
L.J. Slater,  ``Generalized hypergeometric functions,''  
 Cambridge University Press (2008).

\bibitem{Oprisa:2005wu} 
  D.~Oprisa and S.~Stieberger,
``Six gluon open superstring disk amplitude, multiple hypergeometric series and Euler-Zagier sums,''
  hep-th/0509042.
  
\bibitem{Smirnov}
V.A. Smirnov, 
``Evaluating Feynman Integrals,''
Springer Berlin Heidelberg, November 2010. 
  

 \bibitem{StSt} 
  G.~Puhlf\"urst and S.~Stieberger,
``Differential Equations, Associators, and Recurrences for Amplitudes,''
  Nucl.\ Phys.\ B {\bf 902}, 186 (2016)
  [arXiv:1507.01582 [hep-th]].



 \bibitem{Broadhurst:1998rz} 
  D.J.~Broadhurst,
``Massive three - loop Feynman diagrams reducible to SC* primitives of algebras of the sixth root of unity,''
  Eur.\ Phys.\ J.\ C {\bf 8}, 311 (1999)
  [hep-th/9803091].

\bibitem{Kalmykov:2008ge} 
  M.Y.~Kalmykov and B.A.~Kniehl,
``Towards all-order Laurent expansion of generalized hypergeometric functions around rational values of parameters,''
  Nucl.\ Phys.\ B {\bf 809}, 365 (2009)
  [arXiv:0807.0567 [hep-th]].
   
  
   \bibitem{Kotikov:1991pm} 
  A.V.~Kotikov,
``Differential equations method: New technique for massive Feynman diagrams calculation,''
  Phys.\ Lett.\ B {\bf 254}, 158 (1991);
 ``Differential equation method: The Calculation of N point Feynman diagrams,''
  Phys.\ Lett.\ B {\bf 267}, 123 (1991);\\
  E.~Remiddi,
``Differential equations for Feynman graph amplitudes,''
  Nuovo Cim.\ A {\bf 110}, 1435 (1997)
  [hep-th/9711188].
  
 \bibitem{Chetyrkin:1981qh}
  K.G.~Chetyrkin and F.V.~Tkachov,
  ``Integration by Parts: The Algorithm to Calculate beta Functions in 4 Loops,''
  Nucl.\ Phys.\ B {\bf 192}, 159 (1981).



 \bibitem{Argeri:2007up} 
  M.~Argeri and P.~Mastrolia,
  ``Feynman Diagrams and Differential Equations,''
  Int.\ J.\ Mod.\ Phys.\ A {\bf 22}, 4375 (2007)
  [arXiv:0707.4037 [hep-ph]];\\
  J.M.~Henn,
  ``Lectures on differential equations for Feynman integrals,''
  J.\ Phys.\ A {\bf 48}, 153001 (2015)
  [arXiv:1412.2296 [hep-ph]].


 \bibitem{Henn:2013pwa} 
  J.M.~Henn,
  ``Multiloop integrals in dimensional regularization made simple,''
  Phys.\ Rev.\ Lett.\  {\bf 110}, 251601 (2013)
  [arXiv:1304.1806 [hep-th]].


\bibitem{Kalmykov:2006pu}
  M.Y.~Kalmykov,
``Gauss hypergeometric function: Reduction, epsilon-expansion for integer/half-integer parameters and Feynman diagrams,''
  JHEP {\bf 0604} (2006) 056
  [hep-th/0602028].

\bibitem{Kalmykov:2006hu} 
  M.Y.~Kalmykov, B.F.L.~Ward and S.~Yost,
``All order epsilon-expansion of Gauss hypergeometric functions with integer and half/integer values of parameters,''
  JHEP {\bf 0702}, 040 (2007)
  [hep-th/0612240].



  \bibitem{Remiddi} 
  E.~Remiddi and J.A.M.~Vermaseren,
  ``Harmonic polylogarithms,''
  Int.\ J.\ Mod.\ Phys.\ A {\bf 15}, 725 (2000)
  [hep-ph/9905237].
  

\bibitem{Gon2001}
A.B.~Goncharov, 
``Multiple polylogarithms and mixed Tate motives,'' (2001), math.AG/0103059.

\bibitem{Duhr} 
  C.~Duhr,
  ``Hopf algebras, coproducts and symbols: an application to Higgs boson amplitudes,''
  JHEP {\bf 1208}, 043 (2012)
[arXiv:1203.0454 [hep-ph]].

  \bibitem{Kalmykov} 
  M.Y.~Kalmykov, B.F.L.~Ward and S.A.~Yost,
``On the all-order epsilon-expansion of generalized hypergeometric functions with integer values of parameter,''
  JHEP {\bf 0711}, 009 (2007)
  [arXiv:0708.0803 [hep-th]].
  
\bibitem{Ablinger:2015tua} 
  J.~Ablinger, A.~Behring, J.~Bl\"umlein, A.~De Freitas, A.~von Manteuffel and C.~Schneider,
``Calculating Three Loop Ladder and V-Topologies for Massive Operator Matrix Elements by Computer Algebra,''
  arXiv:1509.08324 [hep-ph].


  \bibitem{Boels} 
  R.H.~Boels,
``On the field theory expansion of superstring five point amplitudes,''
  Nucl.\ Phys.\ B {\bf 876}, 215 (2013)
  [arXiv:1304.7918 [hep-th]].

\bibitem{Goncharov} A.B. Goncharov, 
``Multiple $\zeta$-values, hyperlogarithms and mixed Tate motives,''
preprint, 1993. 



\bibitem{LD}  J.A. Lappo-Danilevskij, ``M\'emoires sur la thŽorie des 
syst\'emes des \'equations 
diff\'erentielles lin\'eaires. Vol. IIIÓ, Travaux Inst. Physico-Math. Stekloff, 8, Acad. Sci. USSR, Moscow--Leningrad, 1936, 5--206.

\bibitem{Panzer}
  E.~Panzer,
``Algorithms for the symbolic integration of hyperlogarithms with applications to Feynman integrals,''
Comput.\ Phys.\ Commun.\  {\bf 188}, 148 (2014).
[arXiv:1403.3385 [hep-th]].
  
 \bibitem{BrownFuchs}
F. Brown, 
``Single-valued hyperlogarithms and unipotent differential equations,''  preprint, 2004.

\bibitem{Davy}
E.E.~Boos and A.I.~Davydychev,
``A method of evaluating massive Feynman integrals,''
Theor.\ Math.\ Phys. {\bf 89}, 1052 (1991)

\bibitem{Davydychev:2000na} 
  A.I.~Davydychev and M.Y.~Kalmykov,
  ``New results for the epsilon expansion of certain one, two and three loop Feynman diagrams,''
  Nucl.\ Phys.\ B {\bf 605}, 266 (2001)
  [hep-th/0012189].
  
\bibitem{Smirnov:2008iw} 
  A.V.~Smirnov,
  ``Algorithm FIRE -- Feynman Integral REduction,''
  JHEP {\bf 0810}, 107 (2008)
  [arXiv:0807.3243 [hep-ph]].
  
\bibitem{Anastasiou:1999ui} 
  C.~Anastasiou, E.W.N.~Glover and C.~Oleari,
``Scalar one loop integrals using the negative dimension approach,''
  Nucl.\ Phys.\ B {\bf 572}, 307 (2000)
  [hep-ph/9907494].
  
  \bibitem{Progress} Work in progress.  


\end{thebibliography}
\end{document}